\documentclass[twocolumn]{article}
\usepackage{amssymb}
\usepackage{amsmath}
\usepackage{amsthm}
\usepackage{authblk}
\usepackage[font=small]{caption}
\usepackage[margin=1in]{geometry}
\usepackage{graphicx}
\usepackage{listings}
\lstdefinelanguage{pseudocode}
{
    keywords=[1]{
        write_kernel, write_algorithm, write_schedule_space,
        create_stripe_config, set_config_params
    },
    keywordstyle=[1]\bfseries,
    keywords=[2]{tune_schedule, select_schedule},
    keywordstyle=[2]\itshape,
    basicstyle=\footnotesize\ttfamily,
}
\usepackage{subcaption}

\newcommand{\defn}{\emph}
\theoremstyle{definition}
\newtheorem{definition}{Definition}

\title{Stripe: Tensor Compilation via the Nested Polyhedral Model}
\author[1]{Tim Zerrell}
\author[1]{Jeremy Bruestle}
\affil[ ]{tim.zerrell@intel.com, jeremy.bruestle@intel.com}
\affil[1]{Intel}

\begin{document}

\maketitle

\begin{abstract}
    Hardware architectures and machine learning (ML) libraries evolve rapidly.
    Traditional compilers often fail to generate high-performance
    code across the spectrum of new hardware 
    offerings. To mitigate, engineers develop hand-tuned kernels for each 
    ML library update and hardware upgrade. Unfortunately, this approach
    requires excessive engineering effort to scale or maintain with any
    degree of state-of-the-art performance. Here we present a Nested Polyhedral
    Model for representing highly parallelizable computations with limited
    dependencies between iterations. This model provides an underlying
    framework for an intermediate representation (IR) called Stripe, amenable
    to standard compiler techniques while naturally modeling key aspects of modern 
    ML computing. Stripe represents parallelism, efficient memory layout, and 
    multiple compute units at a level of abstraction amenable to automatic 
    optimization. We describe how Stripe enables a compiler for ML in the style
    of LLVM that allows independent development of algorithms, optimizations, 
    and hardware accelerators. We also discuss the design exploration advantages of 
    Stripe over kernel libraries and schedule-based or schedule-space-based code 
    generation.
\end{abstract}

\section{Introduction}
    \label{sec-intro}
    
    The industry is producing an explosion of varied and creative hardware 
    accelerator architectures \cite{eyeriss-v2-18,shidiannao-15,eie-16,tpu-17,
    moons-accel-17,scnn-17,yin-accel-17}. Designs tend to be optimized for 
    specific goals, such as power efficiency or performance, and often breed 
    even more specialized architectures for very targeted use cases, such as 
    the MAERI \cite{maeri-18}. To take advantage of these innovative designs,
    ML software must be optimized to target the specialized hardware.
    
    \subsection{Kernel Libraries}
        Algorithmic developments aimed at improving accuracy, 
        training or inference performance, regularization, and more continue 
        to progress rapidly \cite{swish-17, groupnorm-18}.
        Nevertheless, these typically retain the same regularities that make 
        specialized ML architectures feasible, and could in principle be 
        efficiently run on ML accelerators. However, the traditional kernel 
        library approach requires a kernel for each hardware-network 
        architectural feature pair, creating a combinatorial explosion of
        optimization work that is infeasible with the rapid growth of both 
        hardware and algorithm designs.
        
        One way to achieve all these optimizations is to write an extensively
        customized kernel to account for each supported machine learning operation
        and each materially different input and output shape
        on each supported hardware platform. As new hardware architectures develop, 
        new kernels must be written for all supported operations. If new operations 
        are devised, they must be added to each hardware target. This coupling 
        creates a maintenance nightmare; decoupling them via automatic code generation,
        where the hardware configuration is separate from both the optimization
        passes and the operations themselves, would be ideal. Unfortunately,
        general-purpose compilers are not aware of the regularities of ML workloads 
        (as discussed in section \ref{sec-requirements}), and so the required 
        optimizations are intractable.
    
    \subsection{Compilers}
        LLVM \cite{llvm-02} is a community-driven compiler infrastructure that
        demonstrates
        features we want in a compiler for ML workloads. Its suite of optimization
        and transformation passes applied to a consistent IR makes it possible to
        optimize workloads in a progressive and modular fashion. Direct transformation
        of the IR allows for deeper 
        changes than those available to a model maintaining some fixed structure 
        along with a modifiable implementation or interpretation. Moreover, an 
        excellent community of experts with varying priorities actively contributes 
        to the ecosystem. Unfortunately, and despite these benefits, LLVM is still
        general purpose and cannot be directly used to compile high performance ML code.
        
        Special-purpose optimization and compilation techniques have also been
        developed. Loop nest transformation and compilation algorithms
        \cite{kennedy-mckinley-93, lim-lam-99}, including the development of the
        polyhedral model \cite{anderson-lam-93, wolf-lam-91}, 
        optimize constrained loop nests via tiling and data layout transformations.
        Frameworks for extracting fine-grained parallelism in traditional workloads
        and applying such polyhedral techniques, including Polly \cite{polly-12} and
        PLuTo \cite{pluto-08}, have proven beneficial. However, these techniques have
        not been sufficient to achieve peak performance for many ML workloads
        \cite{tiramisu-19}.
            
        Various frameworks, including URUK \cite{girbal-06}, Halide
        \cite{halide-13}, and Tiramisu \cite{tiramisu-19} separate loop nest
        semantics from execution order via a scheduling language. TVM \cite{tvm-18} also does 
        this, building on Halide by creating a tensor expression language and 
        adopting a ``decoupled'' scheduling approach that allows for 
        hardware-specific optimizations. The result is a cleaner
        separation of expertise between network architecture and hardware 
        design; see for example Liu et al.~\cite{liu-tvm-cpu-18} on optimizing for CPUs in TVM. 
        AutoTVM \cite{autotvm-18} introduces automatic selection of a schedule from a schedule search space 
        using a deep learning approach with transfer learning. 
        This means hardware-specific optimizations can be written with a focus on 
        getting the right structure (whether to try tiling, whether to try
        different memory layouts (and which), whether to try tensorization,\ etc.)
        without needing to manually experiment with the exact parameters for these
        optimizations. These schedule spaces are still coupled to both the operation
        and the hardware architecture.
        
        Relay \cite{relay-18} is an IR for tensor expressions used in the TVM stack. 
        While its functionality has some overlap with Stripe (transformations enabling 
        tensorization, for example), it is mostly a higher level IR than Stripe; many 
        of its tasks are represented in Tile in the PlaidML stack (automatic 
        differentiation for example) or even at the graph level.
        
        nGraph \cite{ngraph-18} provides optimization opportunities at the graph level,
        where the network-to-device compilation can be managed with a series of
        ``subgraphs''. Since 
        graphs can be managed at this level in both a static and dynamic manner, the 
        performance increase can be used to further accelerate training workloads, 
        or (as is the more common use case for nGraph) to output inference computations 
        in environments where low-latency is important. nGraph may be used in conjunction
        with PlaidML (see section \ref{stripe-in-plaidml}) to provide complementary
        graph optimizations.
        
        Glow \cite{glow-18} offers graph compilation and does not generate code for
        operations like GEMMs or convolutions, instead relying on kernel libraries or 
        accelerator-specific compilers.
        
        Other machine learning domain specific compilers include XLA \cite{xla-17},
        Diesel \cite{diesel-18}, DLVM \cite{dlvm-17}, and
        TensorComprehensions \cite{tensor-comprehensions-18}.
    
    \subsection{Stripe}
        We propose a compiler structured along the same lines as LLVM: it lowers source 
        code to an intermediate representation (IR) and selects and parameterizes a list 
        of optimization passes from a common pool; these passes are then iteratively 
        applied to the IR; and only after all have been applied is the IR code lowered 
        to hardware-specific instructions. A key innovation of our proposed compiler is
        the IR, 
        called Stripe, which abstracts to a granularity fine enough to represent the full 
        new functionality available on ML accelerators and coarse enough to allow
        automatic compilation of high performance ML code.
        
        \begin{figure*}[tp]
            \centering
            \begin{minipage}{0.33\textwidth}
                \centering
                \textbf{Kernel Library}
                \begin{lstlisting}[
                    basicstyle=\footnotesize\ttfamily,
                    keywordstyle=\bfseries,
                    morekeywords={write_kernel},
                ]
foreach HW Architecture
    foreach HW Version
        foreach Kernel     
            foreach Input Shape
                foreach Output Shape
                    write_kernel
                    
                    
                \end{lstlisting}
            \end{minipage}%
            \begin{minipage}{0.33\textwidth}
                \centering
                \textbf{Schedule Space Searching}
                \begin{lstlisting}[
                    language=pseudocode,
                ]
  foreach Kernel
      write_algorithm
      foreach HW Architecture
          write_schedule_space
          foreach HW Version
              foreach Input Shape
                  foreach Output Shape
                      select_schedule
    			\end{lstlisting}
   			 \end{minipage}%
    		\begin{minipage}{0.33\textwidth}
    			\centering
    			\textbf{Stripe}
    			\begin{lstlisting}[
                    basicstyle=\footnotesize\ttfamily,
                    keywordstyle=\bfseries,
                    morekeywords={write_algorithm, create_stripe_config, set_config_params},
                ]
  foreach Kernel
      write_algorithm
      
  foreach HW Architecture
      create_stripe_config
      foreach HW Version
          set_config_params
              
                \end{lstlisting}
            \end{minipage}
            \caption{
                A comparison of the manual engineering needed under different code generation approaches. For a schedule search space approach like AutoTVM, autotuning is used to select the schedule; other scheduling approaches may instead manually write schedules for the various hardware versions and tensor shapes (and thus do not use a schedule search space in favor of manual schedule ``selection''). For Stripe, note that hardware configuration is done independently of the kernels.
            }
            \label{3-col-foreach}
        \end{figure*}

        Stripe is built to represent tensor operations via the Nested Polyhedral
        Model (Section \ref{sec-nested-poly}). This model nests polyhedra in the 
        sense that, for each point in a parent polyhedron, a child polyhedron is 
        defined. This nesting naturally represents tiling, partitioning, tensorization, 
        and other ``blocking'' operations. It also allows assignment of nested 
        polyhedra to nested memory units, giving the compiler a way to match 
        the compute structure to the caching structure for multi\-level hardware 
        topologies.

        At the same time, when this hardware-based complexity is not needed,
        Stripe does not require it to be
        specified. Stripe code representing a single tensor operation can be
        represented as an unnested polyhedron, and a network can be represented 
        as a list of polyhedra. This allows straightforward lowering to Stripe from a
        language that uses a syntax directly representing mathematical formulas 
        for the tensor operations (PlaidML's Tile language, for example).
        The representation is then transformed through a series of optimization
        passes to divide and rewrite these basic polyhedra into nested polyhedra
        appropriate for maximizing performance on the hardware target.
            
        We see several key advantages to Stripe. Foremost, Stripe removes the 
        combinatorial explosion of engineering work from the interaction between 
        growth in accelerators and growth in operations. The classic compiler 
        approach of Stripe means that algorithms can be written on a per-operation 
        basis and optimizations can be written on a per-architecture basis; notably, 
        neither must be written based on both the operation and the hardware 
        architecture. Even with a schedule-space autotuning approach like AutoTVM, 
        schedule spaces must be written for each combination of operation type and
        architecture 
        type. For kernel libraries, manually-engineered code must also include 
        hardware and operation parameters (see Figure \ref{3-col-foreach}).
        
        Stripe's compiler provides modular and extensible optimization passes, allowing 
        novel optimizations without requiring redevelopment of existing optimizations. 
        Stripe's optimization passes are generic and parameterized, enabling reuse 
        across any hardware target for which the pass is beneficial. Stripe's nested 
        polyhedral model naturally represents memory hierarchies of nested and 
        potentially-heterogeneous depth, thereby supporting complex hardware 
        topologies. The compilation model of Stripe doesn't require physical
        hardware or even a cycle-accurate model, just a selection of optimization
        passes with appropriate parameters; in contrast to autotuning approaches
        this allows software-hardware codesign early in the development cycle
        and at relatively low cost.
    
\section{Requirements of Current Machine Learning Execution}
    \label{sec-requirements}

    To successfully produce high-performance code, an ML compiler must first
    analyze, accurately, any defining features in the dataflow and then perform 
    tractable optimizations to target complex hardware topologies based on those
    features. Most ML frameworks today that proffer state-of-the-art performance 
    do not have a compiler that satisfies these requirements, and thus instead 
    use expansive kernel libraries.

    \subsection{Data Use Analysis} \label{sec-data-use-analysis}
        Analyzing the performance of a machine learning workload with any degree 
        of accuracy requires clear analysis of data usage. Particularly important
        are \emph{how much} data is used (i.e., Is dataflow split into appropriately sized
        chunks for the memory units being used?) and \emph{which} data is used
        (i.e., What produces and depends on this data? How much reuse is possible
        if the data is retained at various levels of the memory hierarchy?). 
        Tracking details such as these in a general-purpose compiler can be 
        extremely challenging, and even limited solutions are important 
        research areas.
        
        Machine learning workloads are typically highly structured in several 
        ways that provide key clues for how our analysis should proceed. ML 
        workloads have few control dependencies (exceptions are generally minimal 
        and straightforward: reversing the non-padding portion of a sequence in
        a recurrent network may depend on the length of the sequence, for example).
        Thus, we can calculate, rather than estimate, what data will need to be
        used or reused. Moreover, the calculations necessary to track data dependency and
        aliasing for machine learning workloads are often reasonably
        straightforward and tractable. The natural representation of data 
        is typically a tensor with several dimensions of various sizes. The
        operations performed generally access input and output tensors in a
        highly regular manner; specifically, an iteration space of indexes 
        can be defined, and all tensor accesses can be defined as affine 
        polynomials in terms of indexes in the iteration space. Additionally,
        intra-operation dependencies most frequently take the form of
        commutative and associative aggregations, such as the sum of a
        multiply-accumulate or the max of a maxpool.
        
    \subsection{Complex Hardware Topologies}
        Appropriately distributing ML tasks to accelerators requires optimizations 
        not typically required in the CPU or GPU cases, nor are they required at 
        the same level of generality. A GPU may need explicit memory movement 
        between internal compute units, though it is unlikely to need this across
        multiple levels of memory hierarchy or multiple memory partitions. A CPU
        might need vectorization to use vector instructions, but it is unlikely to
        have compute units that operate \emph{only} on tensorized data. Partitioning
        work amongst multiple heterogeneous hardware units may also be necessary
        to appropriately distribute an ML workload.
        
        Supporting even one accelerator will very likely require memory management at 
        multiple levels and distribution of work to compute units with varied 
        capabilities. With a kernel library, utilization of these features will be 
        written directly into the kernel. With a compiler, optimizations appropriate 
        to these features will be automatically generated. An optimization 
        that decides which data should be moved from larger distant memory to smaller 
        closer memory (at whatever level of the overall hierarchy these memory units 
        reside) readily generalizes to multiple accelerator architectures. Similarly, 
        an optimization that can distribute work to heterogeneous compute units of 
        varied complexity will generalize to varied accelerator architectures.
        
        A hardware runtime will still be necessary, even with a compiler. However, 
        with the compiler performing general machine learning optimizations targeted 
        to the hardware architecture, the runtime can be developed in a low-level,
        hardware-facing manner without requiring optimizations for specific ML tasks.
        
    \subsection{Tractable Optimizations}
        The patterns of control flow and data use in machine learning
        workloads make optimization easier, not harder. Many optimizations
        become tractable, and indeed necessary, to achieve state-of-the-art
        performance. In both programmer-managed and automatically-cached memory
        scenarios, data should be loaded in a manner that maximizes reuse and 
        takes full use of available space without spilling -- this usually 
        involves division into multiple and distinct levels of memory. Often 
        there are hardware-specific instructions requiring exact stencil sizes, 
        especially on accelerators. Where multiple hardware units are available, 
        the work must be appropriately balanced, even when the units provide 
        heterogeneous functionality. Finally, complex scheduling problems arise 
        from doing all of this in a context with deep chains of massive and 
        mostly parallel operations.
        
        \paragraph{Autotiling}
            Large tensors may need to be split into smaller tiles to optimize
            cache reuse. Autotiling must evaluate the performance of potential
            tilings and split loops into tiles accordingly. The autotiling pass
            drives many Stripe design choices and will be discussed in further
            detail in Section \ref{sec-autotiling}.
        
        \paragraph{Microarchitectural Transposition}
            Advanced instructions or specialized compute units may require data
            in a specific layout. Code that could take advantage of these
            instructions or compute units if its data were transposed must be
            found, and the transposition performed.
        
        \paragraph{Microarchitectural Stenciling}
            The microarchitecture may need a specific tile size (stencil), in
            addition to the required dimension-order for its data layout. Code
            that could use specialized instructions or compute units if the data
            matched a specific stencil must be found, and that data must be 
            reshaped to the stencil.
        
        \paragraph{Banking and Partitioning}
            It may be useful for multiple compute units to work in parallel on
            different portions of the same data. For operations that can be run
            in parallel in this way, the relevant tensors must be partitioned
            into different compute unit-specific caches or into different banks 
            to enable this parallel work without conflict.
        
        \paragraph{Fusion}
            To maximize cache reuse, it may be better to perform multiple
            operations on only one or a few tiles of data before proceeding to
            other data. Code may include a series of loops that could potentially
            share the same outer loop and internally perform those operations in
            serial. The relative performance of such a fusion must be compared to
            other possible fusions (or no fusion at all); where a fusion is
            valuable, the code must be rewritten to a fused form.
        
        \paragraph{Scalarization and Memory Localization}
            Transient intermediates produced in registers may not need to be
            stored into memory and reloaded into registers. Temporary memory may
            only be needed in inner portions of the memory hierarchy. Memory
            allocation must be pulled inside loops where legal and semantically
            equivalent, and unnecessary stores and loads must be found and
            eliminated.
        
        \paragraph{Scheduling}
            Operations reading and writing logical tensor data must be rewritten
            to access physical device memory.  This requires assigning physical
            memory locations for logical tensor data, scheduling data movement
            to and from the physical memories accessible to compute units, and
            reordering the operations to take advantage of data locality.
            
        \paragraph{Separating Interior and Boundary Tiles}
            Some workloads do not evenly divide into tiles, or they might have 
            special boundary conditions or other irregularities that do not 
            affect most tiles, but that must be considered nonetheless. These 
            irregularities are best handled separately from the general tiles.
    
\section{Stripe Design \& Implementation}
    The Stripe IR is designed to provide both a \emph{level} and \emph{type} of 
    granularity appropriate to optimizing machine learning tasks. In Section 
    \ref{sec-nested-poly} we discuss the Nested Polyhedral Model, which provides 
    the theoretical underpinnings for Stripe. In Section \ref{sec-stripe-struct}
    we describe the Stripe IR, discussing how it implements the Nested Polyhedral
    Model, and how this implementation enables important optimizations. In Section 
    \ref{sec-autotiling} we detail how autotiling is performed when compiling 
    Stripe, and demonstrate how optimization passes function with Stripe.
    
    \subsection{Nested Polyhedral Model} \label{sec-nested-poly}
        \subsubsection{The Polyhedral Model}
    
            \begin{definition}
                An integer polyhedron $\mathcal{P}$ is a set of all
                $\vec{x}\in\mathbb{Q}^n$ such that
                \begin{align*}
                    &A \vec{x} + \vec{b} \geq \vec{0}\text{, and} \\
                    &A \vec{x} + \vec{b} \in \mathbb{Z}^m
                \end{align*}
                where $A\in\mathbb{Q}^{m\times n}$ and $\vec{b}\in\mathbb{Q}^m$.
            \end{definition}
            Note that this definition is not equivalent to the definition
            sometimes used of an integer polyhedron as a set of $\vec{x} \in
            \mathbb{Z}^n$ satisfying $A \vec{x}+ \vec{b}\geq \vec{0}$ (e.g.\ in
            Bondhugula et al.~\cite{pluto-08}); instead, it is the intersection of a lattice with
            a real convex polyhedron. For convenience, we will use the term
            ``polyhedron'' to refer specifically to bounded integer polyhedra
            that are subsets of $\mathbb{Z}^n$.
            
            The polyhedral model \cite{anderson-lam-93, feautrier-92, girbal-06,
            pugh-91, wolf-lam-91}
            is a model for performing iterative
            computations over an index space defined by a polyhedron, with
            dependencies between steps generally also defined.
            This paper will not go into detail on this model; an overview can be
            found in Girbal et al.~\cite{girbal-06}. Instead, we will develop a \defn{nested
            polyhedral model} of iterative computation that most notably differs
            from the polyhedral model in its dependency structure.
            
        \subsubsection{Parallel Polyhedral Blocks}
            
            In the Nested Polyhedral Model, there are no dependencies between
            iterations, with the possible exception of reduction dependencies.
            This is specified more precisely in Definition \ref{defn-block}.
            
            \begin{definition} \label{defn-block}
                A \defn{parallel polyhedral block} $(\mathcal{P},
                \mathcal{S}_\mathcal{P}, \mathcal{D}, \mathcal{A}_\mathcal{D})$
                consists of a polyhedron $\mathcal{P}$ called the
                \defn{iteration space}, a map $\mathcal{S}_\mathcal{P}$ from
                points in $\mathcal{P}$ to lists of \defn{statements}, a set of
                I/O \defn{buffers}
                $\mathcal{D}$, and a map $\mathcal{A}_\mathcal{D}$ from
                buffers of $\mathcal{D}$ to associative and commutative
                operations called the \defn{aggregation operations} satisfying
                the following:
                \begin{enumerate}
                    \item Statements in $\mathcal{S}_\mathcal{P}$ may only read
                    or write to buffers in $\mathcal{D}$ or to internally-scoped
                    temporaries that are not shared between iterations. A single
                    statement list $S_p\in\mathcal{S}_\mathcal{P}$ may have
                    arbitrary dependencies between its statements and is
                    interpreted as running serially.
                    \item If the statements $s_i\in\mathcal{S}_\mathcal{P}$ for
                    iteration $i\in\mathcal{P}$
                    write to a buffer element $b\in B\in\mathcal{D}$, no
                    statements $s_j\in\mathcal{S}_\mathcal{P}$ for
                    $j\in\mathcal{P}$, $j\neq i$ may read from this buffer
                    element $b$.
                    \item When a buffer element $b\in B\in \mathcal{D}$ is
                    written to by statements in statement lists $S_{i_0}, \dotsc,
                    S_{i_n}\in\mathcal{S}_\mathcal{P}$ for multiple index values
                    $i_0, \dotsc, i_n \in \mathcal{P}$, the value written to $b$
                    is
                    \begin{align*}
                        a_B(v_{i_0}, \dotsc, v_{i_n})
                    \end{align*}
                    where $v_{i_0}, \dotsc, v_{i_n}$ are the values for $b$
                    computed by the statement lists $S_{i_0}, \dotsc, S_{i_n}$ and
                    $a_B\in\mathcal{A}_\mathcal{D}$ is the aggregation operation
                    associated with $B$.
                    
                    When a buffer element $b$ is written to by the statements in
                    statement list $S_i$ for exactly one iteration $i\in\mathcal{P}$, then the
                    value $v_i$ computed for element $b$ by $s_i$ is written to
                    $b$, regardless of the aggregation operation.
                \end{enumerate}
            \end{definition}
            
            Imposing these dependency restrictions makes it more straightforward to
            parallelize execution over different elements of the iteration space.
            The only necessary additions to the statements in $\mathcal{S}_\mathcal{P}$
            involve how to handle aggregation (i.e.\ adding aggregation operations and 
            temporary storage for intermediate results---and even this may be 
            unnecessary if, for example, the aggregations can be made atomic). 
            At the same time, the statements within a block are semantically serial 
            (although they may, of course, be executed in parallel if the compiler 
            can determine that doing so is equivalent to serial execution), and thus 
            within a block (intra-block) statements may have complex dependencies 
            (except insofar as they modify the externally-visible buffers in $\mathcal{D}$).
            
            Note that as defined, a parallel polyhedral block need not
            demonstrate the regularities common to machine learning workloads.
            Such blocks can involve statements with complex control dependencies
            (they are restricted in what buffer elements they can read or write
            by other iterations); they make no restrictions requiring affine
            data access patterns; they can have statement lists that are
            altogether unrelated for different iteration indexes. This makes
            verifying that the dependency conditions are satisfied challenging,
            especially for optimization passes that automatically rewrite the
            parallel polyhedral block to code that must be proven semantically
            equivalent. Moreover, utilizing specialized hardware can be
            challenging. For example, if the statements differ between every
            iteration index, utilizing SIMD hardware effectively is essentially 
            impossible. Additional restrictions Stripe makes to match ML workloads 
            to hardware and to make execution efficient are discussed in
            Section \ref{sec-stripe-struct}.
        
        \subsubsection{Nested Polyhedral Model}
        
            The Nested Polyhedral Model is built from parallel polyhedral blocks
            by defining one or more statements of a parallel polyhedral block to
            be the execution of another parallel polyhedral block.
            
\begin{figure}[htb]
                \centering
                \includegraphics[width=3.1in]{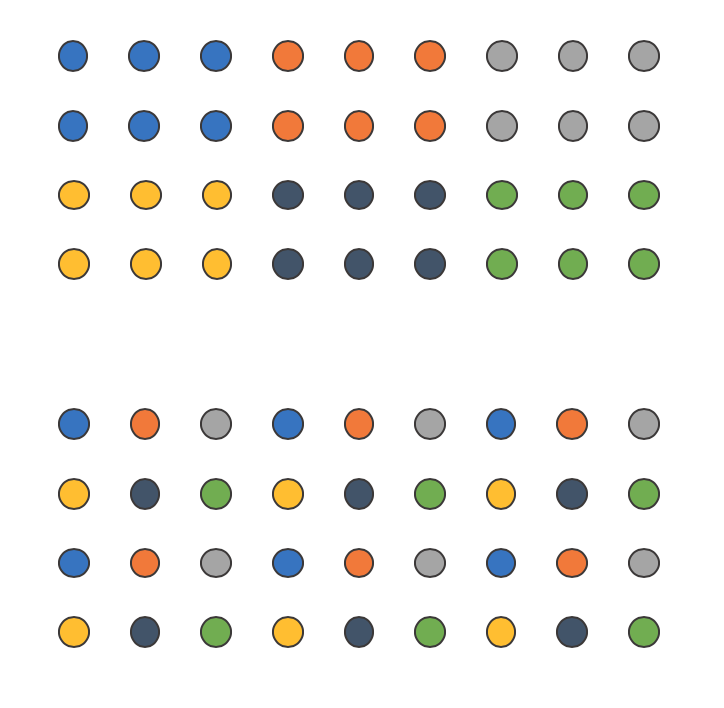}
                \caption{
                    Two tilings of a tensor iterated over by nested polyhedral
                    blocks. Points of the same color are iterated over in the
                    same inner block. In the upper tiling, the inner-block access steps
                    one unit horizontally or vertically as the appropriate index
                    is incremented, while the outer-block access steps three
                    units horizontally or two vertically. In the lower tiling, the
                    outer-block access steps one unit as the appropriate
                    index is incremented and the inner-block access that steps
                    in units of three or two. Either is readily expressed in the
                    Nested Polyhedral Model, and as there are no conflicting
                    accesses, no serial statements need be used. Thus, both are
                    hierarchically parallelizable.
                }
                \label{nested-poly-tiling}
            \end{figure}

            Ensuring that the dependency conditions of Definition
            \ref{defn-block} are satisfied will almost always require the inner
            parallel polyhedral block to depend on the iteration indexes of the
            outer block. Stripe accomplishes this by offsetting
            memory accesses in the inner block based on the iteration indexes of
            the outer block (as well as of the inner block). See Figure
            \ref{nested-poly-tiling} for examples of the resulting access
            patterns; as illustrated, this readily represents ``block'' access
            patterns (such as those arising from vectorization, tensorization,
            tiling, and partitioning).
            
\begin{figure}[htb]
                \centering
                \includegraphics[width=3in]{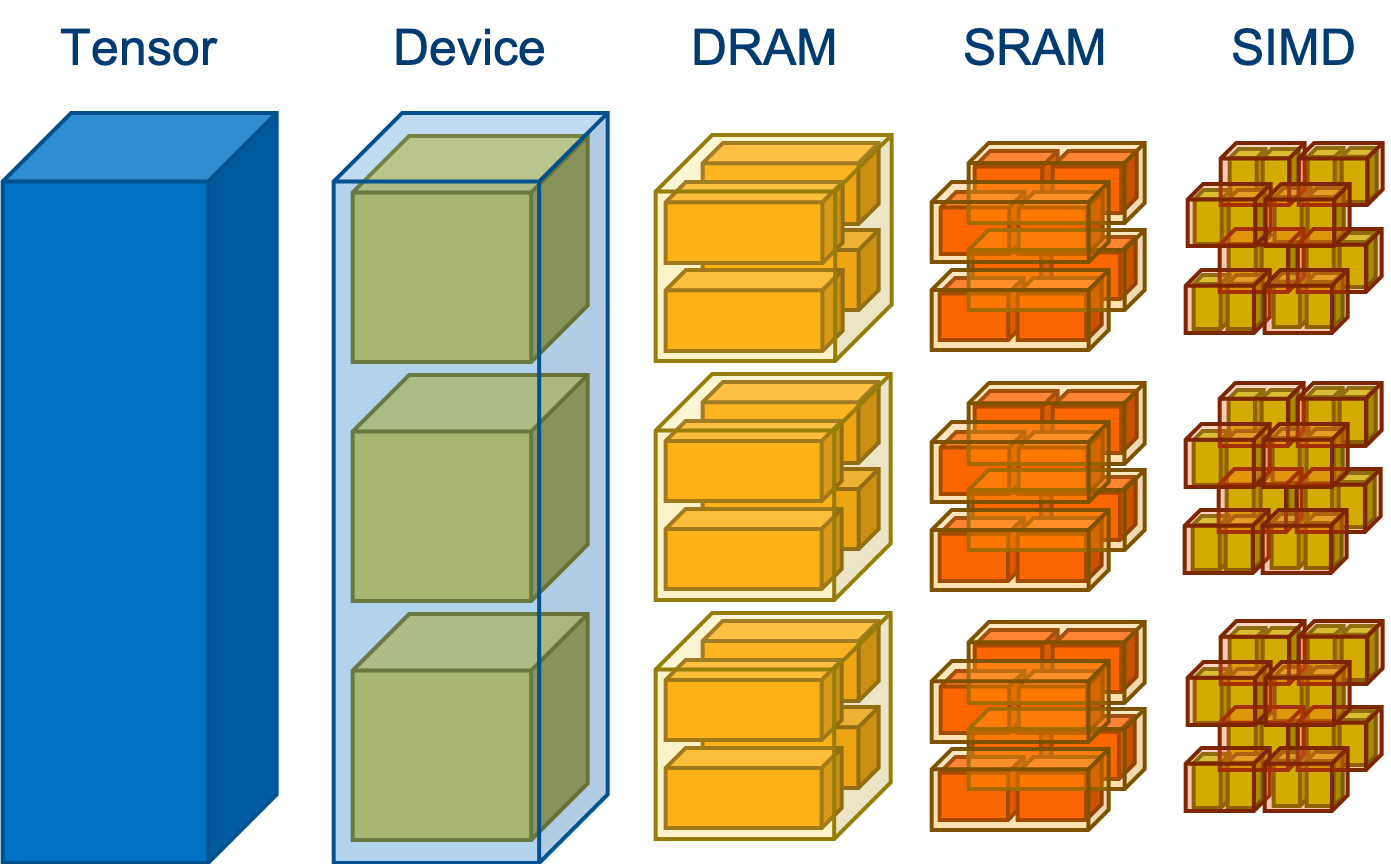}
                \caption{
                    The memory regions accessed by statements in the parallel
                    polyhedral blocks at various levels in a Nested Polyhedral
                    Model. Each column shows the memory accesses from a
                    different nesting depth. The columns are labeled with
                    hardware features that might be targeted by blocks at
                    that level of nesting.
                }
                \label{nested-blocks}
            \end{figure}

            This nesting of parallel polyhedral blocks can be extended to as
            many levels as appropriate for the problem, creating a hierarchy of
            parallelizable code. Figure \ref{nested-blocks} illustrates what
            regions of a tensor might be accessed in a multilevel nest of
            parallel polyhedral blocks constructed from partitioning, tiling,
            and tensorization passes to target a hardware architecture.
            
    \subsection{Structure of Stripe} \label{sec-stripe-struct}
        
        Stripe represents parallel polyhedral blocks with the \defn{block}
        structure. A Stripe block captures the polyhedral iteration space by
        specifying a list of \defn{index} names, a \defn{range} for each index,
        and a list of affine \defn{constraints} on the indexes. There is a single
        \defn{statement list} that does not vary between iteration indexes. The
        statements do access different buffer elements in different iterations,
        and statements that are themselves blocks may have their inner iteration
        space modified based on the outer iteration. With the restriction to
        a single statement list, assigning work to SIMD hardware becomes efficient.
        The I/O \defn{buffers} of a Stripe block are explicitly declared, along
        with an \defn{aggregation} operation for each buffer. Stripe includes an
        \lstinline{assign} aggregation operation that indicates it is illegal
        for values in the buffer to be written to by multiple iterations.
        
        Buffer accesses in Stripe are affine functions of the iteration indexes,
        potentially including indexes of all parent blocks. This makes aliasing
        analysis much easier, which is critical for verifying that all
        properties of a parallel polyhedral block remain satisfied after an
        automatic rewrite. Analysis is also simplified by requiring any parent
        index used to be explicitly passed to the child block.
        
        Stripe statements can be another block, an \defn{intrinsic}, or a
        \defn{special} function. An intrinsic works with scalar values: it can
        read or write a scalar from a buffer (using a buffer access that is an
        affine polynomial of index values as described above), or perform
        simple operations on scalars, such as addition or a trig function.
        Special functions perform complex operations on tensors that are
        inappropriate to represent as blocks of operations on scalars, e.g.\
        scatter or gather.
        
        Operations expressed as scalars in Stripe are not always performed by
        manipulating scalars at the hardware level (e.g.\ due to vectorization).
        For situations where blocks of scalar statements have appropriate
        semantics that translate in whole to hardware instructions, Stripe
        includes \defn{tags} which signal to optimizations passes and the
        lowerer that a chunk of code is intended to be lowered in certain way.
        Tags are more general than just this use case: any element of Stripe
        code may be given an arbitrary set of strings which are its tags. These
        tags have no semantic meaning (in the sense that they do not change the
        expected program output), but instead provide additional information to
        Stripe optimization passes and the hardware abstraction layer. Other use
        cases for tags include storing results from analysis passes to avoid
        repeating the analysis in later passes where such recomputation may be
        expensive or challenging.
        
        To clarify the memory access of a block, all buffers used in a block
        must be explicitly declared, and the scope of a buffer is limited to the
        block it is declared in. In particular, buffers are not in scope within
        child blocks unless explicitly passed to the child. Stripe uses
        \defn{refinements} to declare passing a buffer to a child block. The
        refinement declares whether the child buffer is to be used for input,
        output, or both, and indicates what subregion of the parent block is
        represented---child buffers do not have to bring the entire parent
        buffer into scope in the child block. Typically they don't, which
        enables verification of parallelizability of the nested polyhedral
        structure. A refinement also describes the memory layout of the child
        buffer, indicating the size and stride (i.e.\ memory layout) of each
        dimension. Passing only these restricted views to inner blocks naturally
        represents memory and compute structures for optimizations like tiling
        and vectorization.
        
        Refinements may also include the hardware \defn{location} of the buffer:
        the name of the memory unit (e.g. ``SRAM''), a bank number (if applicable) 
        which may be determined from the iteration indexes if appropriate, and a 
        memory address. Buffer locations are not required, and hardware-specific 
        optimization passes will need to be run before it is possible to set buffer 
        locations sensibly. Specifying buffer locations allows for more precise 
        analysis of available resources and is a crucial step for devices with 
        programmer-controlled memory.
        
        Blocks may contain multiple statements, and these
        statements must be executed as if in serial. However, when the compiler
        can verify that parallel execution would not change the semantics, this
        parallel execution is allowed. A scheduling pass is used on
        multi-statement blocks to construct a directed acyclic graph of
        dependencies between the statements. Where applicable, information about
        the memory access patterns of statements (e.g. from child block
        refinements) is used to determine if statements are independent. This
        can be especially important for partitioning of work into heterogeneous
        units, where distinct block structures are needed for the different
        units.
        
        Stripe allows arbitrary integer polyhedra to be used as the iteration
        spaces of blocks. However, its syntax encourages the use of rectilinear
        constraints by requiring a range to be specified for each index and
        optionally allowing additional non-rectilinear constraints. This structure
        models the almost-rectilinear nature of common operations like
        convolutions with boundary conditions. Maintaining as much rectilinearity
        of constraints as possible in the IR is valuable, as hardware targets
        often perform better on rectilinear iteration spaces but need to handle
        tasks that are not perfectly rectilinear.
        
        One minor divergence of the implementation of Stripe from theoretical
        parallel polyhedral blocks as specified in Definition \ref{defn-block}
        is that aggregation operations may be only \emph{approximately}
        associative and commutative (floating point addition is a common
        aggregation operation that only approximately has these properties, for 
        example). In such situations, executing a Stripe program is ill-defined and
        nondeterministic; however, this nondeterminism typically leads to
        negligible errors in practice for the same reasons floating point errors
        are typically negligible. In situations where this nondeterminism cannot
        be safely ignored, fixed point or integer types may be used instead.
    
    \subsection{Autotiling} \label{sec-autotiling}
    
        \begin{figure*}[p]
            \centering
			\begin{subfigure}[t]{0.45\textwidth}
				\includegraphics[width=3in]{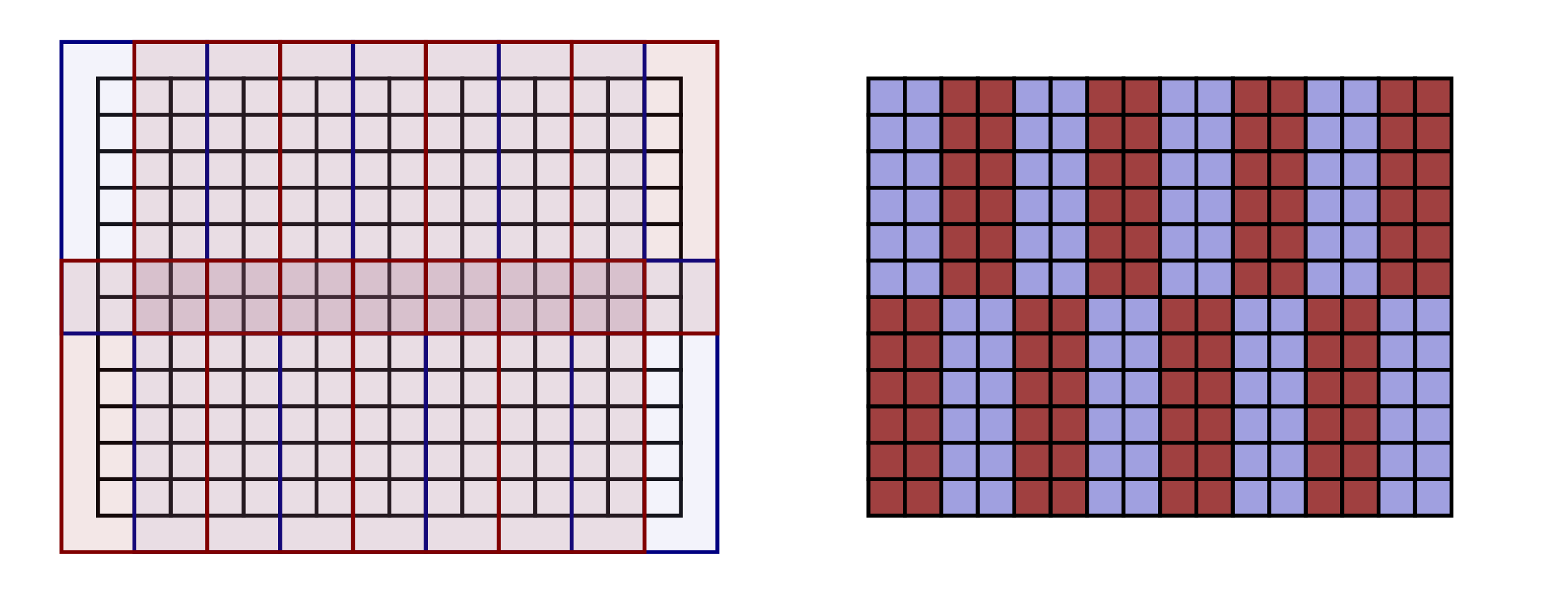}%
				\caption{Cost: 4.666}
			\end{subfigure}\quad%
			\begin{subfigure}[t]{0.45\textwidth}
				\includegraphics[width=3in]{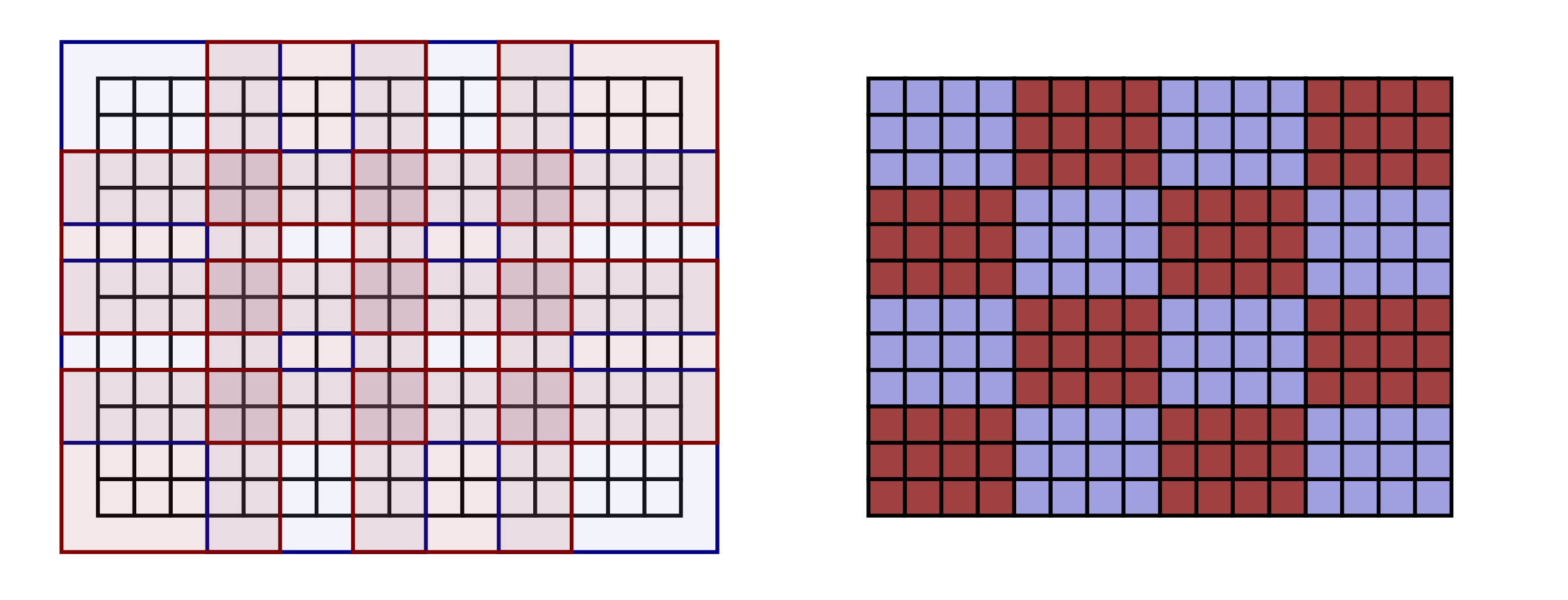}%
				\caption{Cost: 4.5}
				\label{autotile-chosen}
			\end{subfigure}
			\begin{subfigure}[t]{0.45\textwidth}
				\includegraphics[width=3in]{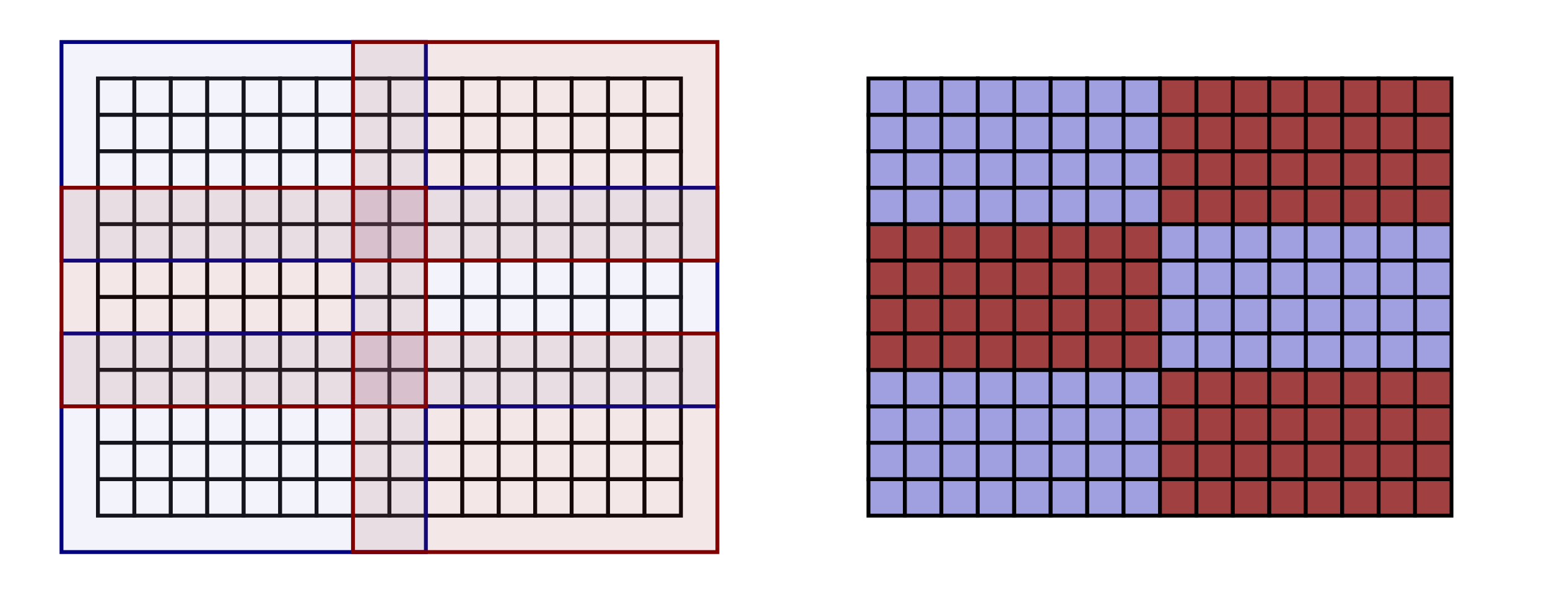}
				\caption{Excluded from search space for requiring too many elements in memory for a single tile.}
			\end{subfigure}\quad%
			\begin{subfigure}[t]{0.45\textwidth}
				\includegraphics[width=3in]{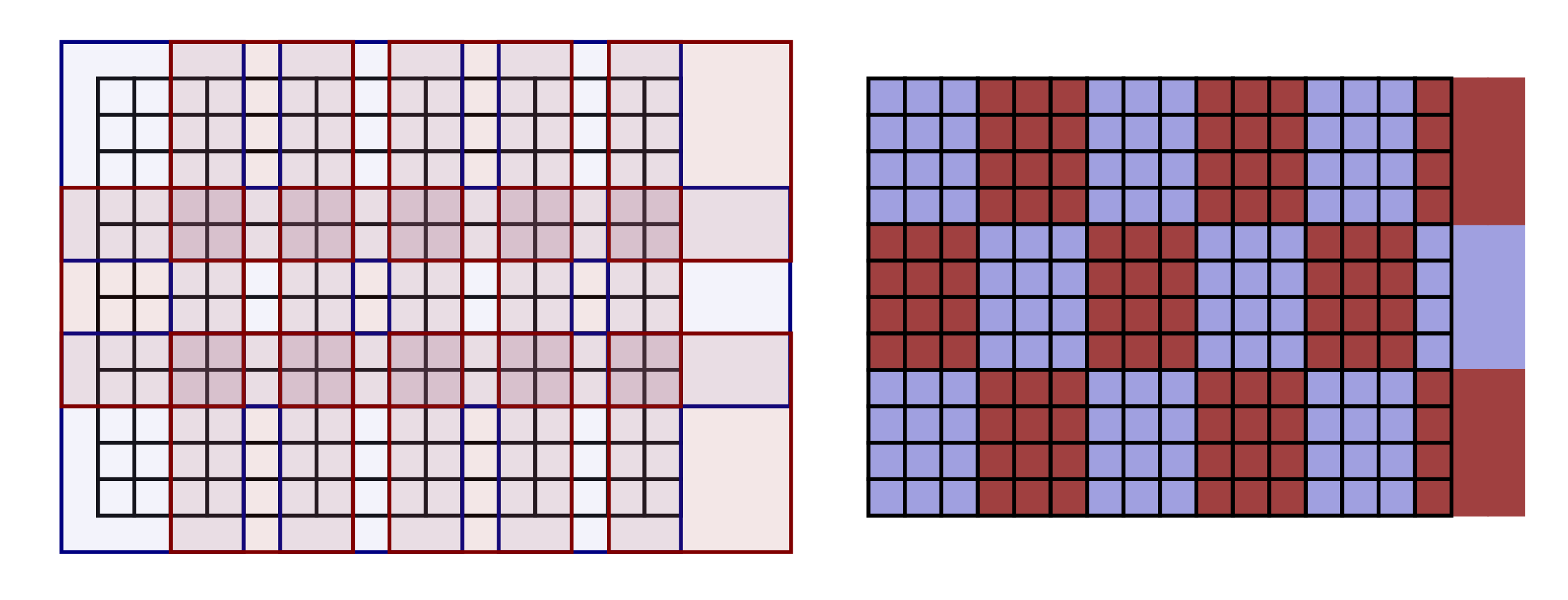}
				\caption{Cost: 5.625.}
			\end{subfigure}
            \caption{Four different tilings along with associated costs. This example shows the input and output tensors of a $3\times 3$ convolution (the weights tensor is not shown and all examples treat it as untiled). We use a hypothetical cost model of number of cache lines accessed, divided by the number of multiply-accumulate operations performed. Tiles on the inputs are shown including overflows; accesses to these elements are removed by constraints in execution but still increase the cost. Only the spatial dimensions are shown, but for the cost model we assume non-batched data with 8 input channels, 16 output channels, and a cache line size of 8 elements. We cap the total memory available for both the input and output tensor tiles at 512 elements.}
            \label{autotile-choices}
        \end{figure*}

\begin{figure*}[p]
	\centering
	\begin{subfigure}[t]{0.5\textwidth}
		\begin{lstlisting}[basicstyle=\tiny]
block []:1 (
	in I[0, 0, 0] i8(12, 16, 8):(128, 8, 1)
	in F[0, 0, 0, 0] i8(3, 3, 16, 8):(384, 128, 8, 1)
	out O[0, 0, 0]:assign i8(12, 16, 16):(256, 16, 1)
) {
	0:
	block [x:12, y:16, i:3, j:3, c:8, k:16] (
		-1 + x + i >= 0
		12 - x - i >= 0
		-1 + y + j >= 0
		12 - y - j >= 0
		in I[x+i-1, y+j-1, c] i8(1, 1, 1):(128, 8, 1)
		in F[i, j, k, c] i8(1, 1, 1, 1):(384, 128, 8, 1)
		out O[x, y, k]:add i8(1, 1, 1):(256, 16, 1)
	) {
		0: $I = load(I)
		1: $F = load(F)
		2: $O = mul($I, $F)
		3: O = store($O)
	}
}		
		
		
		
		
		
		
		\end{lstlisting}
		\caption{Before tiling}
		\label{stripe-before}
	\end{subfigure}%
	\begin{subfigure}[t]{0.5\textwidth}
		\begin{lstlisting}[basicstyle=\tiny]
block []:1 (
	in I[0, 0, 0] i8(12, 16, 8):(128, 8, 1)
	in F[0, 0, 0, 0] i8(3, 3, 16, 8):(384, 128, 8, 1)
	out O[0, 0, 0]:assign i8(12, 16, 16):(256, 16, 1)
) {
	0:
	block [x:4, y:4, i:1, j:1, c:1, k:1] (
		in I[3*x - 1, 4*y - 1, 0] i8(5, 6, 8):(128, 8, 1)
		in F[0, 0, 0, 0] i8(3, 3, 16, 8):(384, 128, 8, 1)
		out O[3*x, 4*y, 0]:add i8(3, 4, 16):(256, 16, 1)
	) {
		0:
		block [x:3, y:4, i:3, j:3, c:8, k:16, xo=3*x, yo=4*y] (
			-1 + xo + x + i >= 0
			12 - xo - x - i >= 0
			-1 + yo + y + j >= 0
			16 - yo - y - j >= 0
			in I[x + i, y + j, c] i8(1, 1, 1):(128, 8, 1)
			in F[i, j, k, c] i8(1, 1, 1, 1):(384, 128, 8, 1)
			out O[x, y, k]:add i8(1, 1, 1):(256, 16, 1)
		)
		{
			$I = load(I)
			$F = load(F)
			$O = mul($I, $F)
			O = store($O)
		}
	}
}
		\end{lstlisting}
		\caption{After tiling}
		\label{stripe-after}
	\end{subfigure}
	\caption{
		Example Stripe code before and after the tiling pass shown in Figure \ref{autotile-chosen}. Note how the iteration space is specified by giving a range for each variable and also specifying any additional non-rectilinear constraints. In Figure \ref{stripe-after}, notice how the overlap between tiles manifests as the size of \lstinline{I} on the middle (tiled) block being larger than the strides of the corresponding spatial indexes. On the innermost block of Figure \ref{stripe-after} the values of the \lstinline{x} and \lstinline{y} indexes on the parent block are explicitly passed in so they may be used in the constraints on the child block.
	}
	\label{autotile-code}
\end{figure*}
    
        To illustrate how Stripe IR automates effective optimizations,
        let's consider one of the key optimization passes: autotiling. Machine
        learning operations are routinely too large and must be split into
        pieces (``tiles'') that fit into local resources.
        The autotiling optimization pass determines the shape of these tiles
        that brings the overall operation's performance closest to the roofline
        \cite{roofline-09} implied by the available compute and I/O bandwidth.
        Depending on the hardware target, several costs may need to be
        considered, including the amount of memory reuse, whether the tile shape
        evenly divides all dimensions of all tensors (and how large any overflow
        is), whether any reductions have been split to multiple tiles and the
        relative cost of computing those reductions later, and the interaction
        of the cache width with the layout of each tensor as restricted by the
        tile shape.
        
        In architectures with automatic caching,
        this tiling optimization improves cache performance by selecting tile
        sizes where full tiles can fit into cache simultaneously, maximizing
        cache hits. In architectures requiring explicit memory transfers, tiling
        determines what data will be transferred, with the tile size ensuring
        that all data fits in the inner memory and that the inner memory is
        efficiently filled to maximize reuse. In architectures with queue-based
        multiprocessing, tiling breaks operations into workgroups effectively.
        
        The autotiling optimization for Stripe explores a space of tile sizes
		using a cost function that models the potential performance impacts
        described above (an example of this is illustrated in Figure
        \ref{autotile-choices}). Several constraints can exclude tile sizes from the
        space to be explored: for instance, the total memory used may not
        exceed the total available memory; also, if the operation is already applied
        to dimensioned blocks (from an earlier vectorization or tensorization
        pass, for example), then the tile size must be an even multiple
        of this size. Search-space heuristics, such as only considering power-of-2 
        dimensions to optionally improve compile performance, may also constrain
        the tile sizes considered.
        
        The design of the Stripe IR makes it straightforward to rewrite blocks
        to introduce an intermediate block of the selected tile size (example code
        produced by such rewriting is provided in Figure \ref{autotile-code}).
        In the
        basic case, simply splitting the index ranges such that the inner iteration 
        space shape matches the selected tile size, and the outer iteration space 
        shape is the quotient of the original index ranges, and passing
        the tensors into the inner block with appropriate offsets will create 
        an effective rewrite. The common complexities of tiling that arise in 
        ML workloads are also readily represented:
        \begin{itemize}
            \item When different coordinates in an output tensor need to read
            from the same coordinates on an input tensor along large dimensions
            (e.g.\ for a non-pointwise convolution), the required iteration
            space
            will be polyhedral and not perfectly rectilinear.
            Constraints representing the boundary / ``halo'' conditions define
            such an iteration space.
            \item For regions that are already not perfectly rectilinear, the
            existing constraints can be pulled into the inner block to maintain
            the same polyhedral structure.
            \item When the optimal tile size does not evenly divide a dimension,
            round up the computed quotient for the outer block (causing an
            overflow). Then remove the overflow by adding a constraint based on
            both the outer and inner index value to not perform any calculations
            in the out-of-bounds overflow region that this introduced.
        \end{itemize}
        
        Stripe's nested block structure readily allows for multiple tiled
        layers. This is useful not only for cases like tensorization, as 
        alluded to above, but also for more general use cases like hardware 
        topologies with multiple layers of memory, or when generally partitioning 
        work amongst multiple units.
        
    \subsection{Stripe in PlaidML}
        \label{stripe-in-plaidml}
        
\begin{figure}[htb]
                \centering
                \includegraphics[width=3.1in]{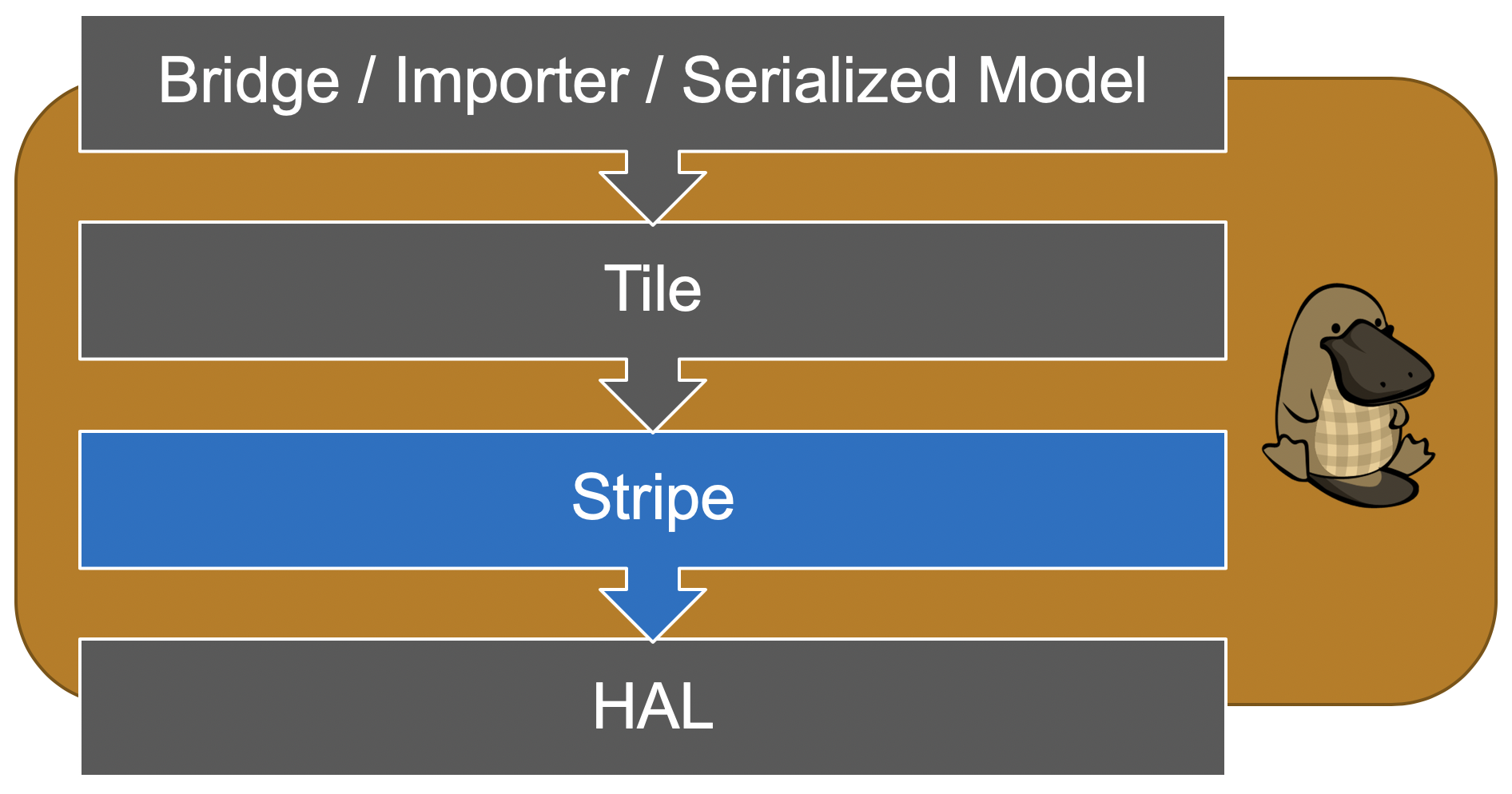}
                \caption{
                    The PlaidML internal stack
                }
                \label{plaidml-stack}
            \end{figure}

        Stripe is the core IR in a larger PlaidML tensor compiler as shown in
        Figure \ref{plaidml-stack}. PlaidML first lowers networks from a source 
        like nGraph \cite{ngraph-18}, Keras \cite{keras-15}, or
        ONNX \cite{onnx-website} into Tile, 
        which is PlaidML's high-level IR representing ML operations in a form 
        reminiscent of Einstein notation. Gradients are computed in Tile if desired, 
        and this Tile code is lowered to Stripe in a general, hardware-agnostic form. 
        Stripe code is then compiled via a series of optimization passes
        targeting the desired hardware, and the resultant code
        is lowered to a hardware abstraction layer, accelerator 
        runtime, or other hardware-appropriate code.
            
\section{Future Work}
    The most crucial future work will be to verify the performance of a variety 
    networks compiled through Stripe on a variety of hardware targets. We are 
    eager to share our approach with the broader community, and we believe that 
    performance for various GPU targets with our pre-Stripe, fixed compilation 
    pass technology demonstrates that Stripe is well-positioned to automatically 
    generate high performance ML kernels. Nonetheless, producing benchmarks for 
    full, modern networks on state-of-the-art hardware is critical. We are 
    actively working on such benchmarks and will be publishing them.
    
    We will also continue to release Stripe on the open source PlaidML GitHub 
    repository. Most notably, while we have released preliminary versions of 
    Stripe already, we do not yet have a release that uses Stripe as part of the 
    main PlaidML compilation pipeline. Producing such a release will be key to 
    making the Stripe IR useful to the open source community.
    
    MLIR \cite{mlir-19} is an upcoming compiler infrastructure providing an IR
    with multiple ``dialects''. This allows IRs of various forms to all be
    embedded as dialects within a broader MLIR infrastructure, improving
    inter-operability and optimization pass reuse. From our current understanding
    of MLIR we believe that both Stripe and MLIR would benefit from adding
    Stripe as an MLIR dialect. In particular, we expect this would provide better
    integration of Stripe with other stages of the machine learning code
    generation and execution pipeline, and would enable greater sharing of
    optimization passes between Stripe and other compilers.

    We hope the extensible nature of Stripe's optimization passes will enable new 
    optimization techniques expressed in the Nested Polyhedral Model to be added 
    to Stripe.
    
\section{Conclusions}
    In this paper, we introduced a domain-specific IR called Stripe that uses 
    the Nested Polyhedral Model to enable automatic generation of machine learning 
    kernels for a variety of hardware targets.  We presented the mathematical 
    underpinnings of the Nested Polyhedral Model, and discussed how it restricts 
    legal schedules that model the extreme parallelism available in machine learning, 
    and how it uses a nesting structure analogous to patterns common in loop nest 
    optimizations and accelerator topologies. We described how a compiler based on 
    Stripe enables powerful, extensible, and configurable optimizations to be 
    developed independently of the machine learning operations and algorithms 
    being optimized.
    
\section{Acknowledgments}
    We would like to gratefully acknowledge the contributions and feedback of a number of people without whom this paper would not have been possible. Particular thanks to Leona Cook for editing, and thanks to Madhur Amilkanthwar, Priya Arora, Mika\"el Bourges-S\'evenier, Cormac Brick, Diego Caballero, Namrata Choudhury, Rob Earhart, Frank Laub, Alessandro Palla, Brian Retford, Mars Saxman, Yao Shi, and Matt Westervelt.

\bibliographystyle{plain}
\bibliography{refs}{}

\end{document}